\documentclass{article}
\usepackage[a4paper]{geometry}

\usepackage[usenames,dvipsnames,svgnames,table]{xcolor} 
\usepackage{microtype}
\usepackage{amsmath}
\usepackage{amsfonts}
\usepackage{comment}
\usepackage{bm}
\usepackage[numbers,sort&compress]{natbib}
\usepackage{enumitem}
\usepackage{mathtools}
\usepackage{tikz}
\usepackage{subfigure}
\usepackage{authblk}

\usepackage{hyperref}
\ifdefined\nohyperref\else\ifdefined\hypersetup
  \definecolor{mydarkblue}{rgb}{0,0.08,0.45}
  \definecolor{mydarkred}{rgb}{0.64,0,0}
  \hypersetup{ %
    pdftitle={},
    pdfauthor={},
    pdfkeywords={},
    pdfborder=0 0 0,
    pdfpagemode=UseNone,
    colorlinks=true,
    linkcolor=mydarkblue,
    citecolor=mydarkred,
    filecolor=cyan,
    urlcolor=magenta,
    pdfview=FitH}

\DeclareMathOperator*{\argmax}{arg\,max}

\title{Human Strategic Steering Improves Performance of Interactive Optimization\footnote{This is the pre-print version. The paper is published in the proceedings of \textit{UMAP 2020} conference. Definitive version DOI: \href{https://doi.org/10.1145/3340631.3394883}{https://doi.org/10.1145/3340631.3394883}.}}

\author{Fabio Colella$^{\dagger1}$}
\author{Pedram Daee$^{\dagger1}$} 
\author{Jussi Jokinen$^{2}$}  
\author{Antti Oulasvirta$^{2}$} 
\author{Samuel Kaski$^{13}$}

\affil{$^1$Helsinki Institute for Information Technology HIIT, Department of Computer Science, Aalto University.  
$^2$Department of Communications and Networking, Aalto University.  
$^3$The University of Manchester. \\
\texttt{firstname.lastname@aalto.fi}\\\smallskip  {\small  $^\dagger$Authors contributed equally.} }

\date{}
\begin{document}

\maketitle
\begin{abstract}

A central concern in an interactive intelligent system is optimization of its actions, to be maximally helpful to its human user. In recommender systems for instance, the action is to choose what to recommend, and the optimization task is to recommend items the user prefers. The optimization is done based on earlier user's feedback (e.g. "likes" and "dislikes"), and the algorithms assume the feedback to be faithful. That is, when the user clicks “like,” they actually prefer the item. We argue that this fundamental assumption can be extensively violated by human users, who are not passive feedback sources. Instead, they are in control, actively steering the system towards their goal. To verify this hypothesis, that humans steer and are able to improve performance by steering, we designed a function optimization task where a human and an optimization algorithm collaborate to find the maximum of a 1-dimensional function. At each iteration, the optimization algorithm queries the user for the value of a hidden function $f$ at a point $x$, and the user, who sees the hidden function, provides an answer about $f(x)$. Our study on 21 participants shows that users who understand how the optimization works, strategically provide biased answers (answers not equal to $f(x)$), which results in the algorithm finding the optimum significantly faster. Our work highlights that next-generation intelligent systems will need user models capable of helping users who steer systems to pursue their goals.

\end{abstract}

\section{Introduction}

\begin{figure*}
\centering
  \includegraphics[width=\textwidth]{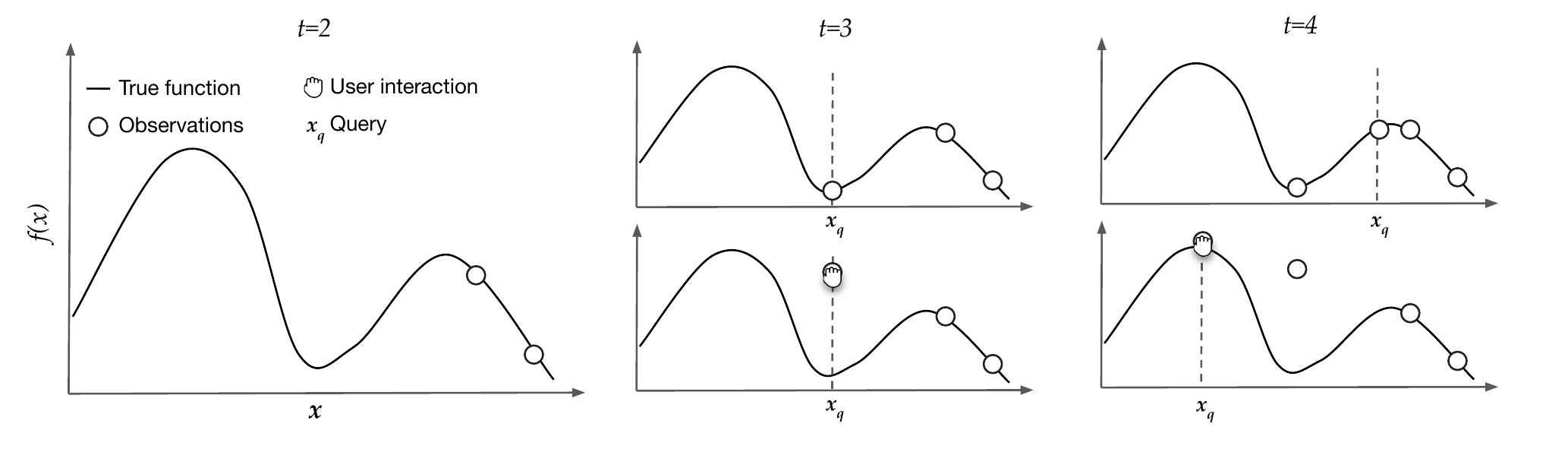}
  \caption{The interactive optimization task studied in this paper. At each iteration, the optimization algorithm queries a point on the $x$ axis, $x_q$, and observes what the user decides to tell about $f(x)$. We show the true function $f$ to the users and allow them to decide on the function values. At iteration $t=3$ if the user answers with the actual function value $f(x)$, the optimization algorithm thinks the optimum is close to the local maximum (top figures). However, if the user gives a biased higher answer ($t=3$ bottom), the algorithm begins to explore on to the left where the uncertainty is high because of the lack of observations.}
  \label{fig:teaser}
\end{figure*}

Interactive intelligent systems with humans in the loop are becoming increasingly widespread. These can range from a personalized recommender system asking about user preference about a recommendation \cite{elahi2016survey, portugal2018use,ruotsalo2015interactive}, or a user guiding the results of a machine learning system \cite{steeringclassification,sacha2017you,amershi2014power,daee2017knowledge,afrabandpey2019human}, to a precision medicine system asking expert's opinion about model characteristics or new data \cite{Holzinger2016,Iiris_precision_medicine}. In all these cases, the intelligent system assumes that the user inputs are faithful responses to the requested query. In other words, users are considered as passive oracles. 

However, results show that even when instructed to be an oracle data provider, users still try to not only provide input on current and past actions, but also to provide guidance on future predictions \cite{amershi2014power}. Furthermore, studies suggest that users attribute mental models to the system they are interacting with \cite{Williams_chi} and are able to predict the behaviour of intelligent systems \cite{tango}. For these reasons we argue that intelligent systems should consider users as active planners. 

A real-life example can be found in interaction with a movie recommendation system. Users can provide a liking or disliking feedback for each movie, which the system then uses to recommend new content. Now, clever users can try to answer in a steering
way (e.g., expressing "like" for a movie they are not interested in) to reach their personal goal of receiving some specific recommendations. For example, a user may not appreciate "The Hobbit: An Unexpected Journey" but may express liking with the intent of receiving more recommendations of fantasy movies similar to Tolkien's. We hence use \textit{steering} to refer to user feedback which is different from the factually true value (i.e., in this case the real grade of appreciation of the movie), and analyse how steering behaviour affects the performance of an intelligent system.

We designed a study to investigate the behaviour of users when interacting with an interactive intelligent system. In particular, we consider the fundamental task of interactive optimization in the form of finding the maximum of a 1-dimensional function. Similar settings have been considered in previous works studying how humans perform optimization and function learning \cite{borji2013bayesian,griffiths2009modeling}. However, we analyse the task from a different angle. In particular, \cite{borji2013bayesian} studied the strategies people use to find the maximum of a hidden function. The users had to sequentially decide on the next point $x$ to be queried about the hidden function (and observing the corresponding $f(x)$) with the goal of finding the maximum with as few queries as possible. The results indicated that users' search strategy is similar to a Bayesian optimization algorithm. Our work is fundamentally different from these, in that in our study a Bayesian optimization algorithm queries the $x$ values, and the user, who sees the hidden function, provides $f(x)$. In other words, the idea is that an AI running an optimization algorithm collaborates with the user.

We hypothesize that users who learn a model of the optimization algorithm, are able to steer it towards their goal. To this end, the next section discusses the optimization problem. The user study setting\footnote{Source code is available at \url{https://github.com/fcole90/interactive_bayesian_optimization}.} is introduced in Section \ref{study}. The paper concludes with a discussion about the results and implications. 

\section{Bayesian Optimization}

Consider the problem of finding the argument that maximizes an objective function in a design space $\mathcal{X}$, i.e., $x^* = \argmax_{x \in \mathcal{X}} f(x)$. Now consider that the function of interest $f(x)$ is unknown and expensive to evaluate, and the only way to gain information about it is to sequentially query it, i.e., ask the function value about a point of interest, such as $x_q \in \mathcal{X}$, and observe the corresponding function value $f(x_q)$ (or in general a noisy version of it). The natural goal for this black box optimization problem could be to find $x^*$ with the minimum number of queries. This problem has been extensively studied in Bayesian optimization (BO) literature \cite{BO_review,frazier2018tutorial} and has been addressed in many applications such as  automatic machine learning (searching in space of machine learning models) \cite{hoffman2014correlation}, design of new materials \cite{frazier2016bayesian,BOSS2019}, reinforcement learning (finding the best action of an agent) \cite{brochu2010tutorial}, and personalized search systems \cite{ruotsalo2015interactive,Daee2016}. 

As the target function is hidden, BO builds a surrogate on the function's observations. The surrogate is usually a Gaussian process (GP) regression model \cite{williams2006gaussian} enabling direct quantification of the uncertainty about the target function. Using the surrogate, the optimizer needs to select the next point to query on the target function. As the query budget is limited, the query algorithm needs to compromise between asking queries that would provide new information about the hidden function (for example in areas that have not been explored), versus exploiting the current best guess about the position of the maximum. This is known as the exploration-exploitation trade-off. Upper confidence bound (UCB) \cite{UCBoriginal} is a well-established query algorithm that in each iteration queries the point which has the maximum value for the sum of the mean and variance of the GP surrogate. Previous works have indicated similarity between human search behaviour and the UCB algorithm \cite{borji2013bayesian,wu2018generalization}.

In our study, the optimization algorithm is a GP-based Bayesian optimization model using UCB for querying new points. We allow the users, who can see the target function, to provide answers to the queries.

\section{User Study} \label{study}

\begin{figure*}
    \centering
    \includegraphics[width=\textwidth]{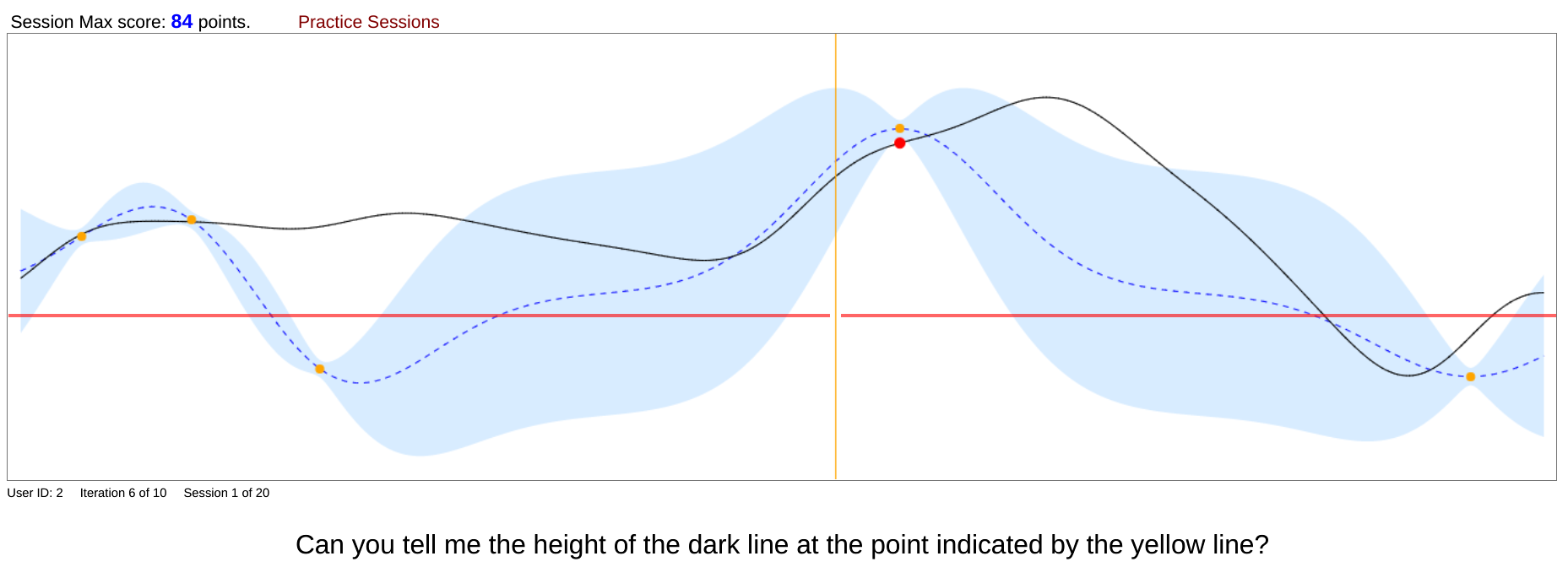}
    \caption{
    The user interface. The true function (solid black) and the surrogate function of the Bayesian optimization (GP mean as dashed blue line and GP confidence bound as light blue band) are visualized to the user. The user responses are shown as orange dots. The query point is indicated by a vertical yellow line while the user can slide the cursor (red line) up and down and then submit their response by clicking. The user score (i.e. the highest point found on the true function) is shown numerically at the top left corner and graphically as a red dot on the true function. The user interface updates after each response.
    }\label{fig:user_interface}
\end{figure*}

\subsection{Method}
\paragraph{Participants}

We recruited $N = 21$ participants for a user study comparing user performance to standard Bayesian optimization in interactive optimization.
The participants were aged 25--35, and there were 12 women. Everyone was awarded one movie ticket upon completion of the study. Eighteen participants had a background in computer science, technology, or engineering, and 16 had a master or a higher academic degree. The participants self-reported their familiarity with GP using a rating scale from 1 to 5.
The mean of the scale was 3.29 ($SD = 1.42$), with 9 participants reporting good knowledge, and 12 poor knowledge.

\paragraph{Materials and procedure}

The experiment consisted of two sessions, with 10 trials of 10 iterations in the first one, and 20 trials of 5 iterations in the second (in addition, both sessions had five practice trials in the beginning).
The session with 10 trials was always conducted first, as it had more iterations and was hence easier, encouraging learning.
Each trial was an independent optimization task, wherein the participant was presented subsequent query points on a randomly generated function. The goal of the participant was to collaborate with the system by providing information about the generated function $f$.
This was accomplished by the system selecting a query point $x$ and by the user selecting a value on the $y$-axis, related to the value of $f(x)$. The mean and confidence bounds of the surrogate function were shown to the user to facilitate the user in learning a mental model of the system. Figure \ref{fig:user_interface} shows a screenshot of the user interface. A questionnaire about background information, including an item about knowledge of Bayesian optimization, was filled in after the optimization tasks, thus avoiding biasing the participants.

\begin{figure}
\centering
    \includegraphics[width=0.6\columnwidth]{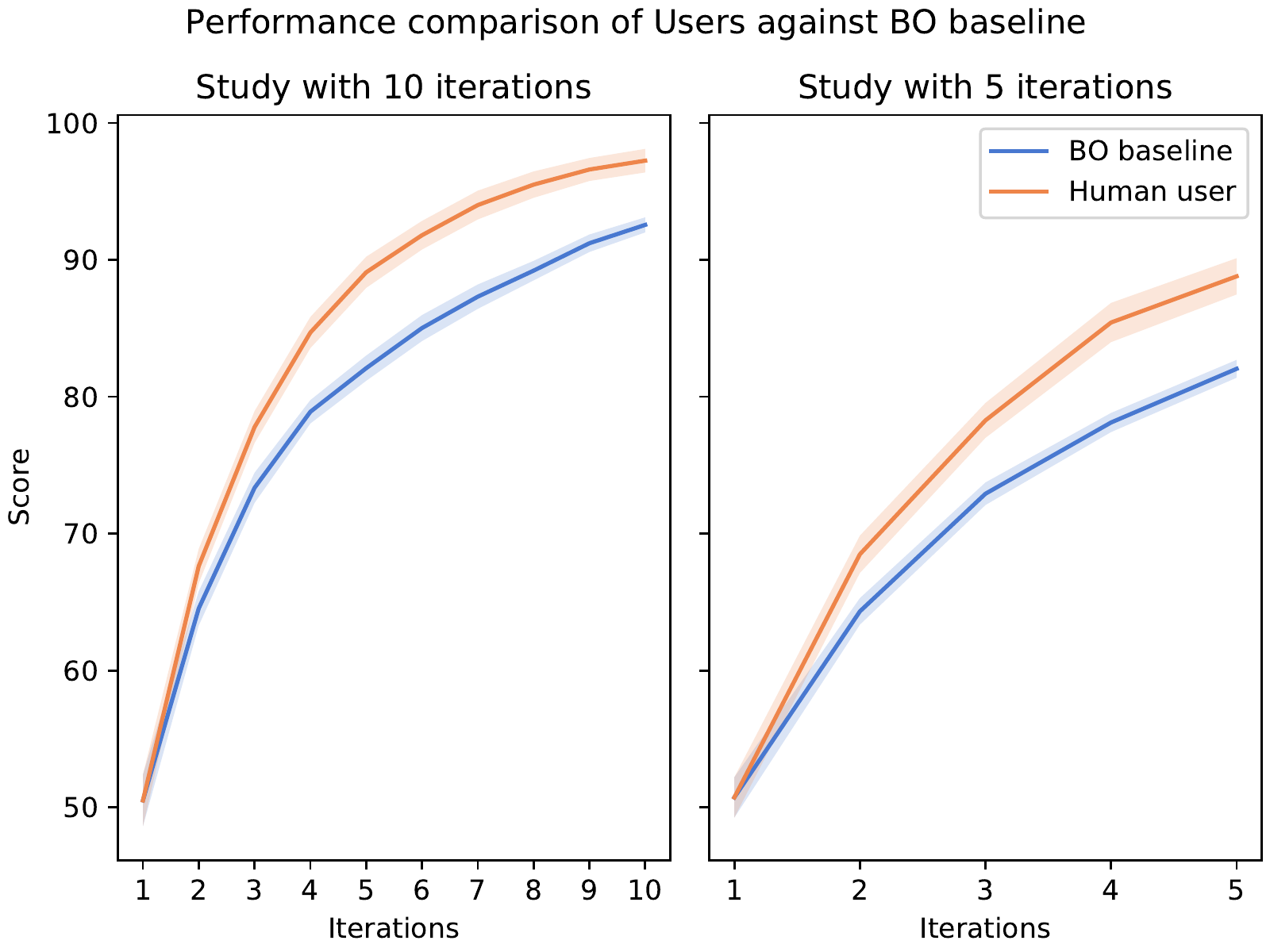}
    \caption{
    Humans performed significantly better than the Bayesian Optimization (BO) baseline. Average score over iterations for all participants (red) and for baseline (blue) for the session with 10 and 5 iterations. The lighter-colored bands around the average lines represent the standard error of the mean.
    }\label{fig:average_scores}
\end{figure}

\paragraph{Data and analysis}
Each response by the participant to the systems' query was considered one iteration. The response was converted into a score reflecting how close the optimization process was to the optimum. The highest function value found so far was reported, normalized between 0 and 100, the score restarted between trials. As baseline we used a standard Bayesian optimization on the same functions that the users optimized, using the same randomized initial query point.
In order to avoid fluctuations due to random effects in the optimization, we averaged the score over 25 runs of the same optimization for each trial.
We tested the following hypotheses, investigating the impact of human collaborator on the score.

\textbf{H1.} Human participants achieve higher scores doing optimization than the baseline Bayesian optimizer.

\textbf{H2.} Human participants achieve higher scores faster than the baseline.

\textbf{H3.} Participants with knowledge of Bayesian optimization perform better than participants without this knowledge.

We tested these hypotheses using mixed regression models with \textit{score} as the dependent variable and \textit{agent} (i.e. human or baseline) as an independent variable. 
H1 was tested on overall performance, aggregated over iterations in each trial.
H2 was tested by adding to the model an interaction effect between \textit{agent} and \textit{iteration}.
For H3, we used only the subset of the data that contained human trials, and compared the performance of knowledgeable (responses to the relevant questionnaire item between 4--5) and naive participants (responses of 3 and below).
Finally, for all tests, we added the participant (\textit{user ID}) as a random intercept into the model.
We report the results using the \texttt{lme4}\cite{lme4} package in R, with Satterthwaite approximations for degrees of freedom.

\subsection{Results}

\begin{figure}
    \centering
    \includegraphics[width=0.6\columnwidth]{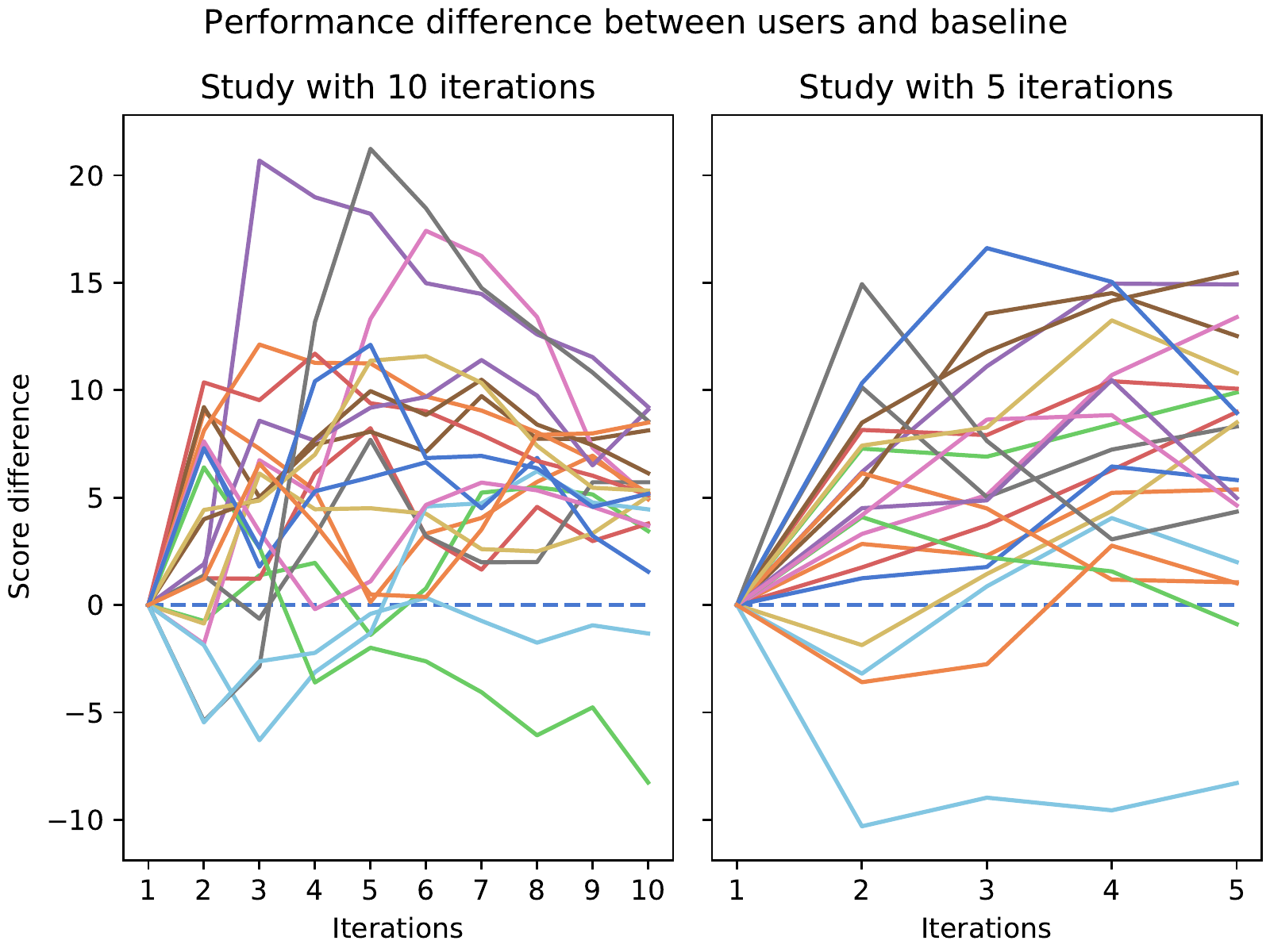}
    \caption{
    Most users performed better than baseline. Each line shows the average difference of the score between one user and the baseline. The difference is computed on the same trial. Value 0 means no difference, while values greater than 0 mean that users performed better than the baseline. We looked at the post questionnaire responses of the two users that performed worse than the baseline (blue and green lines): one reported not understanding the task, while the other mentioned providing random responses in some experimental sessions to explore the system behaviour.
    }\label{fig:improvement_per_user}
\vspace{-0.2mm}
\end{figure}

Compared to the baseline Bayesian optimization, human participants generally performed better, as seen in Figure \ref{fig:average_scores}.
As a trial progresses and iterations increase, humans achieve higher scores than the baseline.
The overall performance was statistically significantly higher for humans (H1), $t(599) = 4.1, p < 0.001$.
Further, human scores increased faster than the baseline (H2), $t(596) = 2.2, p = 0.031$.
Figure \ref{fig:improvement_per_user} illustrates individual score improvement compared to the baseline, by iteration, aggregated over all trials.

\begin{figure}
    \centering
    \includegraphics[width=0.6\columnwidth]{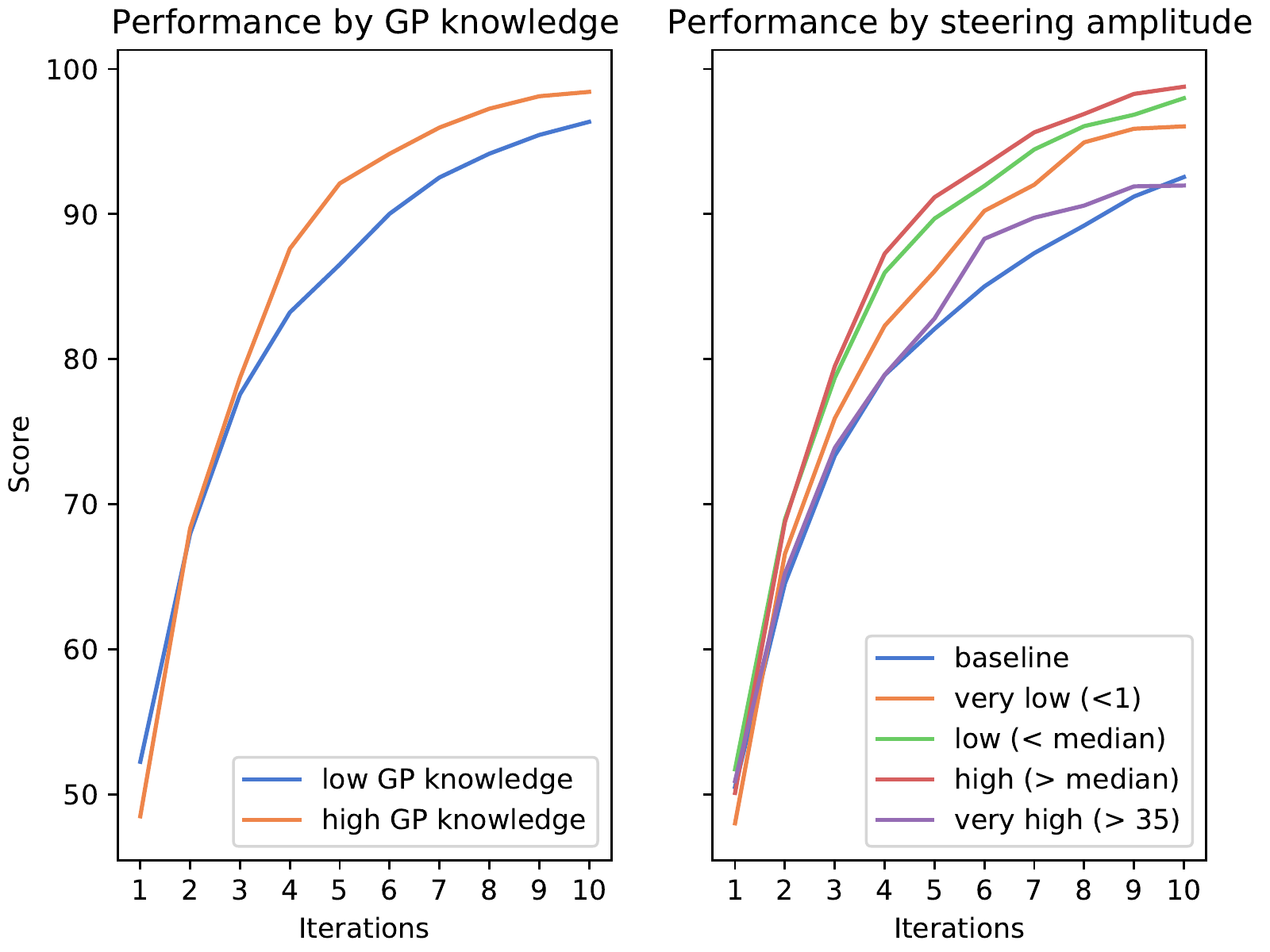}
    \caption{
    Left: users with different understanding of Gaussian Processes have different levels of performance. Right: amplitude of steering influences the average performance. The figure suggests that moderate steering results in improved performance. 
    Similar results were archived in the session with 5 iterations.
    }\label{fig:gp_steer}
\end{figure}

Finally, with the human-only subset of the data, we tested whether users with prior knowledge about Bayesian optimization obtain higher scores than naive users (H3).
Here, the main effect was not statistically significant, $t(51) = -0.4, ns.$, but we did observe a statistically significant interaction effect between iteration and experience, $t(289) = 2.0, p = 0.042$.
This could mean that although both experienced and inexperienced users achieve similar scores in the end, the experienced users can do this with fewer iterations.
This result is shown in Figure \ref{fig:gp_steer}, on the left.

We defined the \textit{steering amplitude} as a value between 0 (no steering at all) to 100 (greatest steering, limited by the user interface height). The average steering was at 19.7 ($SD = 26.13$, $\text{skewness}=1.75$) with most of the user responses far from the actual function values. The performance of the users, grouped by different levels of steering, is shown in Figure \ref{fig:gp_steer}, on the right.

\section{Discussion and Conclusion}

We designed a 1-dimensional function optimization setting to study how humans interact with an interactive intelligent system, here a Bayesian optimization algorithm. Our hypothesis was that while interacting with an intelligent system, humans do not passively provide inputs to the required query but rather they design their input to strategically steer the system toward their own goals. Our results indicate that in case the goals of the human and the algorithm are the same (here finding the maximum of the function faster), human steering behaviour can significantly improve the results. This underlines the importance of developing systems that can understand the mental model of their users \cite{peltola2019teaching, mert2019interactive}. In fact, this strategic behaviour of the user can also leak information about the user's goal which the system could capture to further improve the optimization \cite{brooks2019building}.

Our study was designed with the aim of making the intelligent system's behaviour transparent to the user. For this purpose, we visualized the history of interactions and the algorithm's state (GP mean and its confidence bounds) to the user. In a small pilot we observed that without these elements the users' steering ability was much lower. This suggests that intelligent systems need to be transparent and learnable for the users to take advantage of such steering behaviour. Systems could also infer the user's mental model to render this understanding easier \cite{explanation_soliloquy}.

In conclusion, users strategically steer intelligent systems to control them, and can achieve performance improvements by doing so. This steering behavior could be exploited by the next generation of intelligent systems to further improve performance, but this requires user models capable of anticipating the steering behaviour. Meanwhile, this paper's work could be extended to other application cases such as personalized recommender systems, where the underlying function to be maximized is the user's preference over items.

\subsubsection*{ACKNOWLEDGMENTS}
We thank Mustafa Mert \c{C}elikok, Tomi Peltola, Antti Keurulainen, Petrus Mikkola, Kashyap Todi for helpful discussions. This work was supported by the Academy of Finland (Flagship programme: Finnish Center for Artificial Intelligence, FCAI; grants 310947, 319264, 292334, 313195; project BAD: grant 318559). AO was additionally supported by HumaneAI (761758) and the European Research Council StG project COMPUTED. We acknowledge the computational resources provided by the Aalto Science-IT Project.







\end{document}